\documentclass[prb,preprint,showpacs,superscriptaddress]{revtex4}
\usepackage{graphicx}

\begin{document}
\title{\bf
 CO adsorption on metal surfaces:\\
 a hybrid functional study with plane wave basis set.}

\author{Alessandro Stroppa}
\email{alessandro.stroppa@univie.ac.at}
\affiliation{Faculty of Physics, University of Vienna, and Center for Computational Materials Science,
Universit\"at Wien, Sensengasse 8/12, A-1090 Wien, Austria}

\author{Konstantinos  Termentzidis}
\affiliation{Faculty of Physics, University of Vienna, and Center for Computational Materials Science,
Universit\"at Wien, Sensengasse 8/12, A-1090 Wien, Austria}

\author{Joachim Paier}
 \affiliation{Faculty of Physics, University of Vienna, and Center for Computational Materials Science,
Universit\"at Wien, Sensengasse 8/12, A-1090 Wien, Austria}

\author{Georg Kresse}
\affiliation{Faculty of Physics, University of Vienna, and Center for Computational Materials Science,
Universit\"at Wien, Sensengasse 8/12, A-1090 Wien, Austria}

\author{J\"{u}rgen Hafner}
   \affiliation{Faculty of Physics, University of Vienna, and Center for Computational Materials Science,
Universit\"at Wien, Sensengasse 8/12, A-1090 Wien, Austria}

\date{\today}

\begin{abstract}
 We present a detailed study of the adsorption of CO on Cu, Rh, and Pt (111) surfaces in top and
 hollow sites. The study
 has been performed using the local density approximation, the gradient corrected functional PBE, and the
   hybrid
 Hartree-Fock density functionals PBE0 and HSE03 within
 the framework of generalized Kohn-Sham density functional theory using a plane-wave basis set.
  As expected, the LDA and GGA functionals show a tendency to favor the hollow sites,
 at variance with experimental findings that give the top site as the most stable adsorption site.
 The PBE0 and HSE03 functionals reduce this tendency. In fact, they predict
 the correct adsorption site for Cu and  Rh but fail for Pt.
 But even in this case, the hybrid functional destabilizes
 the hollow site by 50 meV compared to the PBE functional.
  The results of the total energy calculations are presented
  along with an analysis of the projected density of states.

\end{abstract}


\maketitle
\section{Introduction}
The adsorption of carbon monoxide on metal surfaces is an important
case study in surface
science
for two reasons. On one hand, the interaction of CO with
metal surfaces plays a major role in understanding phenomena related to
e.g. catalysis, adhesion and coating as well as in many industrial
processes, such as automotive catalysis, corrosion, tribology, and
gas sensing.\cite{Gabor} Catalysts containing transition metals, such as
rhodium, palladium and platinum have been widely used  to lower the
emissions of CO  in automobile exhausts.\cite{Gabor}  For all
the mentioned applications, it is clear that understanding
adsorption on both bare and adsorbate-covered surfaces is an important
issue.

On the other hand, the failure of Density Functional Theory (DFT)
based on local and semilocal density functionals in predicting the
correct adsorption site for CO on metal surfaces is well known. The
most notable example in literature is the CO/Pt(111) system, often referred to
 as \emph{CO adsorption puzzle}. No "stone was left
unturned"\cite{CO_Pt}
 in order to determine
the reason for the discrepancy between theory and experiment, but neither defect structures
and contaminations, nor
relativistic or spin effects, nor zero-point energies can account for the difference:
there is  strong
evidence that current approximations to DFT underestimate the CO
preference for low-coordination sites.\cite{Kresse1,CO_Pt} Most
plane wave codes
predict that the hollow site is preferred
for CO on Cu, Rh and Pt (with the exception of Ref.~\onlinecite{dacapo} for CO on Rh),
whereas experimentally it is found that CO adsorbs
at the top site with the carbon end down at low coverage on all three
substrates.\cite{expPt1,expPt2,expPt3,expPt4} In all fairness,
it must be emphasized that some local basis set
codes (specifically DMOL and ADF) seem to give the proper site order for
Pt.\cite{spin-orbit,ADF1}
The reason for the discrepancy between local basis set codes and plane
wave codes is not yet entirely understood, but it is likely to be related
to the different treatment of relativistic effects or basis sets. For Pt, the DMOL
code for instance applies effective core potentials to take into
account relativistic effects, and the site preference
depends critically on the used effective core potential, with the most
accurate effective core potential giving the same site preference as plane wave codes.\cite{Dmol1}
The Amsterdam Density Functional code (ADF) also yields the correct site order
for Pt.\cite{ADF1} Basis set convergence, k-point sampling as well as relativistic effects
have been carefully checked for CO on Pt, and it is at this point unclear why results differ
from those reported using plane waves.

Very recently, Q.-M. Hu \emph{et al.}\cite{Scheffler} have shown
that errors of present-day exchange-correlation (xc) functionals are
rather localized and spatially limited
to few nearest neighbors. For extended systems the \emph{correction} can
be estimated  by analyzing properly chosen clusters
 and employing wavefunction based methods for an improved xc treatment. According to their study, this
 procedure applied to CO/Cu(111) and CO/Ag(111) gives the top site as the most stable adsorption site, in agreement with
 the experiments.\cite{Scheffler}
Returning to the discrepancy between theory and experiment obtained using DFT and plane waves, the
 current suggestion  is that the main reason for  the failure is due to the
 incorrect description of the relative position
of the highest occupied molecular orbital (HOMO)  and lowest unoccupied molecular orbital (LUMO) of CO
 with respect to the Fermi energy of the metal.\cite{Gil}
To overcome this problem several possibilities have been considered so far.
 One is to apply an a posteriori correction, based on the singlet-triplet CO excitation
energy  obtained by GGA and configuration
interaction calculations.\cite{way1}
 A second option  is
  using a DFT+U approach, where an additional  U is added to the
DFT Hamiltonian to shift the CO LUMO to higher energies.
\cite{Kresse1,way2} A third self contained and less ad-hoc approach
is the use of hybrid functionals.\cite{Gil}

Hybrid functionals are a combination of exact non-local
orbital-dependent Hartree-Fock (HF) exchange and a standard local
exchange-correlation functional, and they
 provide a significant improvement
over the LDA-GGA description for molecular as well as extended insulating and semiconducting solid
state systems.\cite{Paier} For these systems, hybrid functionals are among the most accurate
functionals available as far as energetics and structural properties are
concerned.\cite{Koch,Scuseria,Scuseria1}


Currently, the most popular ones are PBE0 (or PBE1PBE)
\cite{PBE01,PBE02} and  B3LYP\cite{B3LYP1,B3LYP2}. The
former has been proposed by Perdew, Burke, Ernzerhof,  Adamo and
Barone \cite{PBE03,PBE04} as a ''parameter-free'' functional based
on the PBE exchange-correlation functional. It has promising
performance for all important properties, being competitive with the
most reliable, empirically parametrized  functionals.\cite{PBE02}
The latter was suggested by Becke\cite{B3LYP1} and soon developed
into the most popular and most widely used functional for quantum
chemical calculations.
 This functional
reproduces the thermochemical properties of atoms and
molecules rather well.\cite{Ragha}

For periodic systems, in particular metals, however, the long-range
nature of the Fock exchange interaction and the resultant large
computational requirements present a major drawback. Recently, a new
hybrid functional, called HSE03, has been introduced by J. Heyd
\emph{et al.}\cite{HSE03} This functional addresses this problem by
separating the description of the exchange interaction into a short-
and a long-range part, where the long-range part is treated by
semilocal gradient corrected functionals. The new functional yields
a description of molecular properties comparable to the results
obtained using the PBE0 functional, and, in some cases, it even
gives a slight improvement over the latter.\cite{HSE03}

To the best of our knowledge, very few \emph{ab initio} calculations based on
hybrid functionals have been concerned
 with the problem of CO
 adsorption on
metal surfaces. Gil \emph{et al.}\cite{Gil} reported on the CO
adsorption on the Pt(111) surface using both slabs and cluster
models with local, semilocal and hybrid functionals (B3LYP). But the
B3LYP calculations were restricted to clusters and extrapolation to
large clusters seems to indicate that convergence with respect to
the cluster size was not obtained. The B3LYP functional renders the
on top and fcc sites almost degenerate, whereas LDA and GGA show a
pronounced tendency to favour fcc adsorption. This was confirmed in
the work of Doll \cite{Doll} on the same system,
 where a careful
comparison between gradient corrected functionals and the B3LYP
functional has been reported. It was shown that the B3LYP functional
gives the top site as the preferred site. Finally, in the work of
Neef and Doll,\cite{Neef} the adsorption of CO on the Cu(111) surface has
been studied using the local density approximation, the gradient
corrected functional of Perdew and Wang and the B3LYP functional. The
LDA and GGA yield the fcc site as favorable adsorption site, whereas
 the B3LYP functional results in the preference of the top
site, in agreement with the experiment. The recent study for CO on Cu and Ag(111) come to similar conclusions.\cite{Scheffler}
 All these  hybrid functional studies applied only B3LYP and they made
 use of localized basis sets, which are possibly affected by basis set superposition errors (BSSE).


 The aim of the present report is  to present an  extensive density functional
 study of the adsorption of CO
 on close-packed (111) metallic surfaces using  the PBE0 and HSE03 functionals, which have not been
 considered yet for this specific problem. We also include, for comparison,
 the local density approximation and the standard
  gradient corrected PBE functional,
 which is widely accepted as the best parameter-free density functional available.
 We discuss in detail the application of these functionals to  bulk, bare Cu, Rh and Pt surfaces and the corresponding
 CO adsorption problem.


 The study is pursued  within the framework of the
 plane-wave projector-augmented-wave (PAW) formalism. Based on the approach of Chawla and Voth\cite{Voth}
 for the evaluation
 of the exact exchange, PBE0 and HSE03 functionals have been recently implemented in the Vienna ab--initio Simulation
 Package (VASP).\cite{vasp}
We remark that the use of a plane-wave basis set for the evaluation
of the exact exchange energy
 allows for calculations that are free of basis-set superposition errors and benefit from the computational
 efficiency of fast Fourier transforms. Details of the implementation are given elsewhere.\cite{Paier,Paier2}
  We will focus on three
  reference metal systems, Cu, Rh, and Pt, which are among
  the best studied metallic surfaces concerning CO
  adsorption.\cite{Gabor} We note that the adsorption of CO on Rh(111) has not been investigated
  using hybrid functionals.

   The paper is organized as follows. In the next Section we describe the computational approach and the
model systems used;
  in Sect.~\ref{bulk} we briefly review the calculated properties for the bulk and corresponding bare (111) surfaces as well
  as for the CO molecule; in Sect.~\ref{ads:strut&ene} we discuss the results concerning the structural properties
  and energetics of the adsorbed CO molecule on the surface; Sect.~\ref{ads:ele&charge}
  is devoted to the electronic properties in terms of density of states; in
  Sect.~\ref{Discussion} we discuss the main results of this work and
  in Sect.~\ref{conclu} we draw our conclusions.

\section{Theoretical and computational method}\label{model}
The first principles density functional theory calculations have been
performed within the local and generalized gradient density approximation to
DFT in the Ceperley-Alder and Perdew-Burke-Ernzerhof parametrization, respectively.\cite{LDA,GGA}
 In the present calculations, the interaction between the ions and valence electrons is described by the projector
 augmented wave (PAW)\cite{PAW} method in the implementation of Kresse and Joubert.\cite{K_J}



 The cutoff energy has been fixed to 400 eV, which is sufficient to give well converged
 results for the systems considered in this work.\cite{Paier2}
 The surfaces have been modelled by a periodic four layer metal slab
 with a CO molecule adsorbed on one side of the slab,
 vertical to the (111) surface plane (asymmetric setup).
  For CO on Pt(111), tests with a six layer metal  slab  using the HSE03 functional have also been performed in order to
 check the convergence with respect to the slab thickness: the relative stability of the different sites
 does not change. Each slab is separated from its periodic image in the $z-$direction
 by a vacuum space of $\sim$ 10 \AA.
  The two uppermost surface layers and the CO molecule are allowed to relax (substrate
 buckling up to the second layer below the surface). For the
 electronic relaxation, we fix the energy threshold to 10$^{-3}$ eV whereas, the ionic relaxation is stopped
 when all forces are smaller than 0.1 eV/\AA.
  We use a $c(2\times4)$ in-plane periodicity, which is equivalent to a coverage of $\Theta=0.25$ ML.
  With this choice,
  the separation between the molecule and its in-plane periodic images is large enough
  to neglect spurious adsorbate-adsorbate interactions with a reasonable computational effort.\cite{Hafner}

  For the HSE03 calculations, the range separation parameter $\omega$ was set to $\omega=0.3 \ $\AA$^{-1}$
  in both the density
  functional part as well as the non-local Fock exchange.\cite{ErrorReport}


The Brillouin zone integrations are performed
 on  symmetry reduced grids using the Monkhorst-Pack scheme.\cite{MK}
 To accelerate k-point convergence,
 we set the Methfessel-Paxton\cite{Paxton} smearing
width  to 0.1 eV. We have carefully checked the k-point convergence by performing
 calculations using 4$\times$4$\times$1, 6$\times$6$\times$1
  and 8$\times$8$\times$1 points (see discussion in Sect.~\ref{ads:strut&ene}).
  The numerical accuracy of the adsorption
 energies is estimated
 to be $\sim$ 20 meV.
 We have also investigated  the effect of  downsampling  the reciprocal space representation
 of the Fock exchange operator
 for  the HSE03 functional as outlined in Ref.~\onlinecite{Paier2}. In particular, the Fock exchange
 operator has been evaluated on the ''full'' $n$$\times$$n$$\times$$1$  k-point mesh, as well as, on the
 corresponding ''downsampled''  $\frac{n}{2}$$\times$$\frac{n}{2}$$\times1$  grid, with $n$=4, 6, 8
 (for details see Ref.~\onlinecite{Paier2}). Results are discussed in Sect.~\ref{ads:strut&ene}. Finally,
  we have also tested, for a reference case (HSE03),
  the effect of the parameter that controls the fast Fourier transformation (FFT) grid for the HF routines (keyword ENCUTFOCK in the
  VASP code). By considering the smallest possible FFT grid that just encloses the cutoff sphere corresponding to the
  plane wave cutoff (ENCUTFOCK=0) or simply removing this constraint (without the flag ENCUTFOCK), total energies
  change by less than 5 meV.\cite{VASPmanual}

   We considered three adsorption sites:
   top site, where CO is sitting vertically
   above a metal atom in the top layer, and
 the hcp (fcc) hollow site, where CO is sitting vertically above a
  metal atom in the second (third) topmost layer.

\section{Bulk, bare surface and CO molecule}\label{bulk}

Before studying the effect of CO adsorption on the (111) metal
surfaces, we first consider the bulk and bare  surfaces. In
Table~\ref{TAB1} we summarize the results for the bulk systems
reporting the lattice constant ($a_{0}$) and the bulk modulus ($B_{0}$)
  obtained using the Murnaghan equation of state.
   The cohesive energy ($E_{coh}$) is
   calculated considering unconstrained, i.e. spin polarized and nonspherical,
   ground states of the atoms.\cite{Paier} The numbers between
round brackets give the relative errors with respect to experiment.

 As usual,
the LDA (GGA)  lattice constants are slightly underestimated (overestimated) with respect to
 the experiment. Note that the
 experimental lattice constants reported in
Tab.~\ref{TAB1}  have not been extrapolated to 0 K and
 the zero-point quantum fluctuations have not been included in the
DFT calculations. However, the inclusion of these contributions
should change the   values by only approximately $\sim$
0.1 \%, which  we can neglect for the purpose of the present
study.\cite{Thermal1}

Using hybrid functionals,
the lattice constant is almost unchanged for Cu
 (only a slight expansion is found using PBE0 and HSE03) and
  the overestimation is generally reduced in the other two cases.
For Rh, the hybrid functionals overcorrect
  the lattice constant giving a theoretical value smaller than the
   experimental one; for Pt, it remains slightly larger than
  experiments.
 Except for Cu, the agreement with experiment improves using hybrid functionals (the PBE0 and HSE03
 relative errors are smaller than the  PBE ones).


The overestimation (underestimation) of the lattice constants goes
in hand  with an underestimation (overestimation) of the
bulk-moduli.
The PBE functional predicts
$B_{0}$ slightly more accurately than hybrid functionals
 for the systems under study.

 The cohesive energies are overestimated at the LDA level, they are very well
 reproduced for the
three  bulk metals at the PBE level, but they are  underestimated
using
  hybrid functionals. The underbinding  has been attributed
  to the admixture of Fock exchange, and
 to the fact that Hartree-Fock usually underbinds, especially for metals.\cite{Doll,under1}
 We believe that this is not generally true for hybrid functionals. In fact,
 as shown in Ref.~\onlinecite{Paier2},
  the atomization energies for systems without $d$ electrons, like Li, Na and Al are quite comparable for
   the PBE, PBE0 and HSE03
  functionals (and very close to experimental results within 0.05-0.15 eV). Most probably,
  the reason for the reduction of the
  atomization energies of $d$ metals is related to the increased stability of the
   spin polarized atom compared with an artificial
  non-spin-polarized atom. The hybrid functional overestimates the exchange splitting
  in $d$ elements, with a consequent increase of
  the spin-polarization energy. This however does not explain the discrepancy for Cu.
   Here we believe that the
  neglect of dynamical correlation effects between closed $d$ shells is a major source of errors accounting for
  roughly 210 meV of the discrepancy.\cite{VanderWalls} Proper inclusion of these Van der Waals-like forces
  would also decrease the theoretical lattice constant of Cu.

  For Cu and Pt, we can compare our hybrid functional results with Ref.~\onlinecite{Doll} and
  Ref.~\onlinecite{Neef},
 respectively. From Tab.~\ref{TAB1}, it is evident
  that our results are significantly closer to the experiment
 than those of Ref.~\onlinecite{Doll} and Ref.~\onlinecite{Neef} at the bulk level.
  This is most likely related to the B3LYP
  functional being specifically  designed for small molecules,
  limiting or reducing its precision for heavy elements.\cite{B3LYPPaier}

 In Tab.~\ref{TAB2} we show the relevant properties of the bare surfaces:
 $d_{12}$ is the interlayer distance between the first and
 second   topmost surface layer
 (and the corresponding relative variation with respect to the theoretical bulk value in round
 brackets)
 and the surface energies ($E_{surf}$).

 For all  functionals,
 the Cu and Rh surface layer relaxes inwards.
 For Pt, the top layer relaxes outward. This "anomalous" relaxation  is  already documented
 in  literature.\cite{Wan,Ptexp} The  change of $d_{12}$  relative to the theoretical interlayer bulk
 distance ($\Delta d_{12}$) for Cu is $-$0.9 \% using PBE,
 in  agreement
 with calculations of Neef (PW91);
 the same quantity calculated using PBE0 (HSE03) is $-$1.8 \%
 ($-$0.9 \%) which is again close to the one obtained using B3LYP in
  the same work ($-$1.2 \%).
 For Rh,  $\Delta d_{12}$  is $-$1.6\% (PBE),
 $-$0.6 \% (PBE0), $-$1.0 \% (HSE03)  and for Pt it is
 $+$0.7 \% (PBE), $+$1.3 \% (PBE0) and $+$1.3 \% (HSE03).
 Our   values obtained both using standard GGA and hybrid functionals
 are  slightly smaller than those obtained by Doll.
  The comparison with experiment is not simple due to the
  large uncertainties of the experimental
 values: for Cu $\Delta d_{12}$ is
 $-$1.0$\pm$0.4 \%,\cite{Cuexp}
 for Rh it is $-$1.3$\pm$0.9 \%,\cite{Rhexp} and  for Pt it is $+$1.1$\pm$0.4 \% .\cite{Ptexp}
 Agreement between our calculations and experimental results is certainly reasonable.

 The calculated LDA  surface energies   compare well with previous calculations.\cite{LDAsurf}
 The GGA   surface energy is 0.45, 0.81 and 0.62
 eV, for Cu, Rh and Pt,
 respectively, also in  agreement with the values reported in
 the literature (0.50 eV, 0.81, 0.65).\cite{CuEne,Luckas,Doll}
  The PBE0 and HSE03 functionals give a slightly increased surface energy
  compared to the PBE functional.  Such an increase of the surface
  energies
   is pointing towards an improved description using
 the hybrid functionals, but they are still underestimated compared to experiments:
 the experimental surface energies
  are $\sim$ 0.65 eV for Cu(111), $\sim$ 1.08 eV for
  Rh(111), $\sim$ 1.08  for Pt.\cite{Expsurfene}
    The B3LYP results\cite{Neef,Doll} do not follow a
 consistent trend, with an increased surface energy for Cu  and a decreased one for Pt.
  We have also calculated the  PBE and HSE03 work function for the three (111) metal
  surfaces (not shown in Tab.~\ref{TAB2}).
  For Cu, we have 6.34 (6.10) eV; for Rh, 5.38 (5.05) eV; for Pt, 5.69 (5.64) eV using the PBE (HSE03) functional.
  The experimental values are 4.98, 4.98, 5.65 eV for Cu, Rh and Pt.\cite{workfunction}
  We note that PBE values are overestimated compared to
  experimental values; the HSE03 functional reduces the
  overestimation and gives  better agreement with experiments especially for Rh, and
  Pt.



 Finally for the CO molecule, the calculated bond-length
 ($d_{\textrm{CO}}$) is 1.135, 1.143, 1.133, 1.131 \AA \ for LDA, GGA (PBE), PBE0, HSE03,
  close to the values reported by  Neef (1.150 and 1.140 \AA \ using PW91 and B3LYP\cite{Neef}).
 The experimental  bond length is 1.128 \AA.\cite{dCO}
As far as the HOMO-LUMO gap is concerned we find 6.80, 6.90, 10.60, 8.80 eV
using LDA, GGA (PBE), PBE0, and HSE03 respectively. The hybrid
functionals give an energy gap increased  by $\sim$ 2-3 eV due to
the downshift of the HOMO and a simultaneous upshift of the LUMO.
 The negative of the calculated HOMO energy is 9.0, 8.6, 10.6, 10.0 eV using LDA, GGA (PBE), PBE0 and HSE03 respectively,
  compared to the experimental
 ionization potential of 14.10 eV.\cite{molecules}

\section{CO adsorption: energetics and structural properties}\label{ads:strut&ene}

In Table~\ref{TAB3},~\ref{TAB4} and \ref{TAB5}
 we show the results for the adsorption of a CO molecule on Cu, Rh and Pt surfaces, respectively.
 As expected, using the standard LDA and GGA  functionals
the wrong site is preferred (shown in boldface),
 namely fcc, hcp and fcc for Cu, Rh and Pt in agreement with previous
   calculations.\cite{Doll,Lukas,Neef}
    The order of the sites with respect to the energy (starting from the most stable one)
 is fcc, hcp, top for Cu; hcp, fcc, top for Rh; fcc, hcp and  top for Pt.
 These site orders generally agree with those reported in
 Refs.~\onlinecite{Doll},~\onlinecite{Neef} and \onlinecite{Lukas}, although  different
 surface reconstructions and computational methods were used.
   It should be noted that, for the Cu case, our calculated adsorption energy for
    the most stable site is lower
    by more than 400  meV
   compared to that given in Ref.~\onlinecite{Neef}.
   Also the surface energy
   (bare surface) differs (see previous Section).
    These discrepancies can be possibly related to the
   different computational method used in Ref.~\onlinecite{Neef}, where a
   local basis set has been applied.

   Using  PBE0 or HSE03, the top site is preferred for Cu and Rh,
    in agreement with  experiment.
   For Cu and PBE0 and HSE03, the site order  is top, fcc, hcp  (the  same as found
   in Ref.~\onlinecite{Neef} using B3LYP). The calculated adsorption energy (top site) is about
   $-$0.60 eV.
   This value is also close to that reported in Ref.~\onlinecite{Neef} ($-$0.57 eV).
   The relative energy splitting ($\Delta E$) between hcp and top is $\sim$ 40  meV using PBE0 and $\sim$ 30 meV
   using HSE03. A larger energy splitting (150 meV) was  found by Neef,\cite{Neef} and an even larger one ($\sim$ 200 meV)
    in Ref.~\onlinecite{Scheffler} using B3LYP.
   Furthermore, we find an almost degeneracy between the top and fcc sites using HSE03
   ($\Delta E$=$\sim$6 meV).

   For Rh, the site order   is top, hcp and  fcc using both  hybrid functionals. The calculated
   adsorption energy is $-$2.109 and $-$2.012 eV using PBE0 and HSE03. The splitting top-fcc is only
   $\sim$ 5 meV
   using PBE0 and $\sim$ 16 meV using HSE03, hence certainly within the error bars of the present calculations.
   Therefore, even though the calculated adsorption energies  predict the correct  site order for  Cu and Rh,
   we can only safely  conclude that the applied hybrid functionals
   reduce the tendency of LDA and GGA functionals to  favor the hollow site
   with respect to the top sites.

   Finally, for Pt, neither  PBE0 nor HSE03 recover  the correct site preference:
   the top site remains unfavoured with respect to the fcc site,
   and the top and hcp sites are almost degenerate within the numerical
   accuracy. We, however, note that the top-fcc energy difference is $\sim$ 350 meV (LDA), $\sim$
    160 meV (GGA), $\sim$ 56 meV (PBE0) and $\sim$ 70 meV (HSE03). Therefore,
   the tendency to favour the hollow sites is again reduced using hybrid functionals, but not sufficiently
   so for Pt.
   In order to rule out possible errors due to the k-point mesh,
   we performed additional calculations for the HSE03 case and CO on Pt using
   a 12$\times$12$\times$1 k-point grid and downsampling the HF
   exchange part to  6$\times$6$\times$1 k-points.
   We found that the relative stability
   of the top and fcc sites changed by only $\sim$ 20 meV. Also increasing
  the number of layers to 6 did not change the site order. The presented numbers are thus essentially converged.

    Finally we compare   the calculated adsorption energies  of Cu, Rh and Pt with experimental values.
   For Cu(111), the experimental values are in the range of $-$0.52 to $-$0.46 eV,
   \cite{expadscu1,expadscu2} close to our calculated value ($\approx -$0.60 eV).
   For Rh(111), they are between $-$1.65 and $-$1.43 eV
   \cite{expadsrh1,expadsrh2}. In this case, the theoretical values are too large by $\sim$ 0.40-0.60 eV
   and, worse, the hybrid functionals give a slight increase of the adsorption energy compared to the gradient
   corrected functional.
   This result
   is certainly disappointing, and we will return to it in the conclusions.
  For Pt(111), experimental values are in the range
   of $-$1.71 to $-$1.43 eV,\cite{expadspt1,expadspt2,expadspt3} which compare
   reasonably well with the PBE value. Again the hybrid
   functionals clearly worsen the
   agreement with experiment.

     Concerning the  geometry relaxations, we summarize the trend common to all the three functionals:
      i) the CO bond length is slightly elongated with respect to the  theoretical value for the isolated
   molecule ($d_{\textrm{CO}}$=1.14 ~\AA)
   in the top site, and
   even more so in the hollow sites. This holds for all three metal
   surfaces. ii) The elongation of $d_{\textrm{CO}}$  correlates
   with a corresponding contraction of $d_{\textrm{C-X}}$ from low coordination (top)
   to high coordination sites (fcc, hcp).
    iii)
   The buckling $b$ is larger for the top site.
    iv) $\Delta d_{12}$ varies significantly from top to hcp sites, especially for Cu,
   where  an inward relaxation for atop adsorption  and  an outward relaxation  for the hollow sites
   is observed, whereas an outward relaxation is found  in all three sites  for Rh and Pt.

  We conclude this section with a brief discussion of the k-point convergence
presented in Tab.\ref{TAB2.5}.
Concentrating first on the PBE functional, we note that k-point convergence
is slowest for the Cu surface, but fairly fast for the other two systems.
A relative precision of 10 meV can be attained using 6$\times$6$\times$1
k-points for all three systems [sampling specified for $c(2\times4)$ supercell].
The 4$\times$4$\times$1 k-point set results in errors between
80 meV (Cu), 30 meV (Rh)  and 10 meV (Pt).
The HSE03 functional shows a similar convergence rate, with  errors
being only slightly larger. Now 6$\times$6$\times$1 k-points are sufficient
to yield energies to within a precision of 15 meV, whereas
the 4$\times$4$\times$1 k-point grid causes errors between
120 meV (Cu) and 30 meV (Rh and Pt).
Reducing the k-point sampling for the presentation of the non-local Fock
exchange part has only a very small effect on the relative energies (10 meV), which
is more than acceptable, in particular, in view of a speed up by a factor four if
the sampling for the non-local part of the Hamiltonian
is reduced in $x$ and $y$ direction by a factor 2.

The PBE0 functional, however, converges exceedingly slowly, with discrepancies
between 6$\times$6$\times$1 and 8$\times$8$\times$1 k-points being up to 50 meV.
We were unable to increase the k-point grid beyond 8$\times$8$\times$1 points with our
available computational resources, but the  relative energies using 8$\times$8$\times$1
k-points are typically within 10-20 meV of those obtained using the HSE03 functional,
and we expect them to come even closer to the HSE03 results, if the k-point set
were further improved.
This clearly demonstrates that the HSE03 functional is vastly superior
in terms of computational requirements, in particular, for metallic systems.
The calculations using the HSE03 functional and 6$\times$6$\times$1 k-points with the
Hartree-Fock part presented on a 3$\times$3$\times$1 k-point grid are typically a factor 12 faster
than the PBE0 calculations using 8$\times$8$\times$1 k-points, although both yield
practically the same results.

\section{Electronic properties}\label{ads:ele&charge}

 According to the Blyholder model\cite{Bly}
 the interaction of the CO molecule with a
 transition metal surface
 is usually described as the sum of two contributions. The first is
 due to the overlap of the   highest occupied molecular orbital (HOMO)
 $5\sigma$ with metal states.
 Since this interaction is
 accompanied by donation of electrons from the $5\sigma$ orbital
 into empty metal surface orbitals, this term is called donating
 term. The second term is due to the interaction of the surface-electron
 bands with the   lowest unoccupied molecular orbitals (LUMO) $2\pi^{*}$.
 This is a back-donating term, since now electrons are transferred
 from the metal surface orbitals into the $2\pi^{*}$
 orbitals.  This
 simplistic model has been refined by A.
 F\"{o}hlisch,\cite{Fohlish,PRLCO} using x-ray emission spectroscopy
 and \emph{ab initio} cluster calculations. They showed that
 the $\pi$ bonding is manifested through the creation of a $d_{\pi}$ complex
 related to a hybridization of 1$\pi$ and
 2$\pi^{*}$ orbitals with metal states. The 4$\sigma$ and 5$\sigma$ orbitals and the metal states
 form a hybrid  $d_{\sigma}$ band. The work of  F\"{o}hlisch
 concentrates mainly on the interaction of the frontier orbitals with metal $d$ states,
 however, another often overlooked  issue is that CO chemisorption to a
 transition-metal surface also involves interactions between the
 broad metal $sp$ valence electron band (which contains approximately one electron per atom for Cu,  Rh, Pt)
 and the CO orbitals. Due to symmetry, the $2\pi^{*}$ orbital can not interact with
 the $s$ or $p_z$ orbitals for top site adsorption, but the $2\pi^{*}$-$sp_z$ interaction is
 strong  at high coordination sites, since \emph{antisymmetric}
 combinations of surface $s$ orbitals are available at these sites.\cite{Santen4}

\begin{figure}[!hbp]
\begin{center}

\includegraphics[scale=0.6,angle=-90]{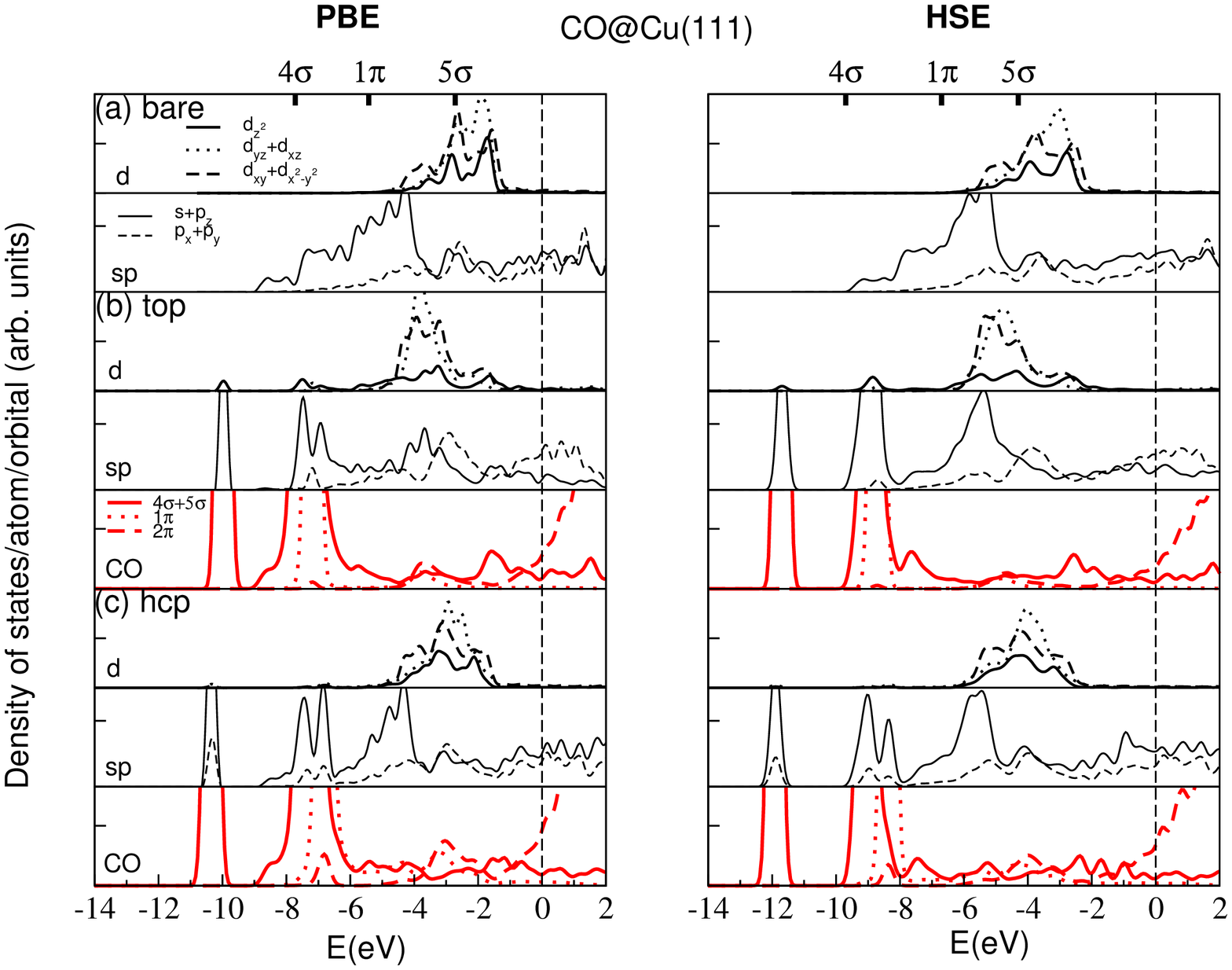}
\\Fig. 1
\end{center}

\caption{(Color online) Orbital resolved electronic densities of states (DOS) for the
topmost Cu layer of the bare surface [panels (a)-d, (a)-sp] and for CO
adsorbed at the top [panels (b)-d, (b)-sp, (b)-CO] and hcp  hollow sites
[panels (c)-d, (c)-sp, (c)-CO]. The DOS are projected
onto the metal $d$ states [panel (a)-d, (b)-d, (c)-d: continuous,
 dotted and dashed thick lines correspond
 to $d_{z^{2}}$, $d_{yz}+d_{xz}$ and  $d_{xy}+d_{x^{2}-y^{2}}$ DOS respectively],
onto metal $sp$ states [panel (a)-sp, (b)-sp, (c)-sp: continuous,
 dashed thin lines correspond to $s+p_{z}$ and $p_{x}+p_{y}$ DOS respectively], and onto molecular
 orbitals [panel  (b)-CO, (c)-CO: continuous,
 dotted and dashed thick red lines correspond to $4\sigma+5\sigma$, $1\pi$, $2\pi^{*}$ DOS respectively].
 The Fermi level
is located at 0 eV. In panel (a)-d, we show the one-electron energies of the $4\sigma$, $1\pi$,
  $5\sigma$
  molecular levels aligned with the Fermi level using bold tick marks.}\label{dosCu}
\end{figure}

\begin{figure}[!hbp]

\begin{center}
\includegraphics[scale=0.6,angle=-90]{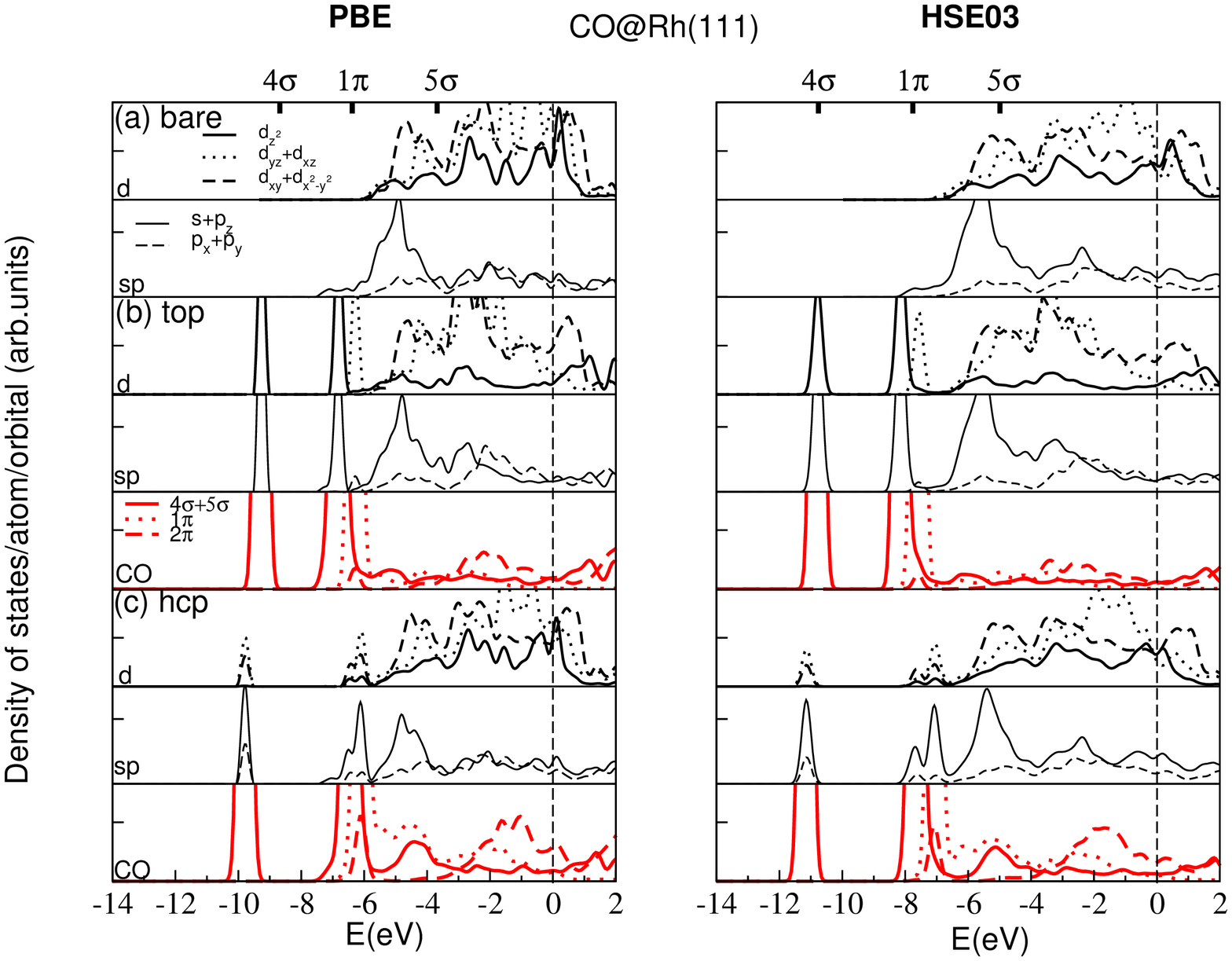}
\\Fig. 2
\end{center}
\caption{(Color online) Orbital resolved electronic density of states (DOS) for the
topmost Rh layer of the bare surface (a) and for CO
adsorbed at the (b) top and hcp (c) hollow site. See Caption of Fig.~\ref{dosCu} for details.}\label{dosRh}

\end{figure}

\begin{figure}[!hbp]

\begin{center}
\includegraphics[scale=0.6,angle=-90]{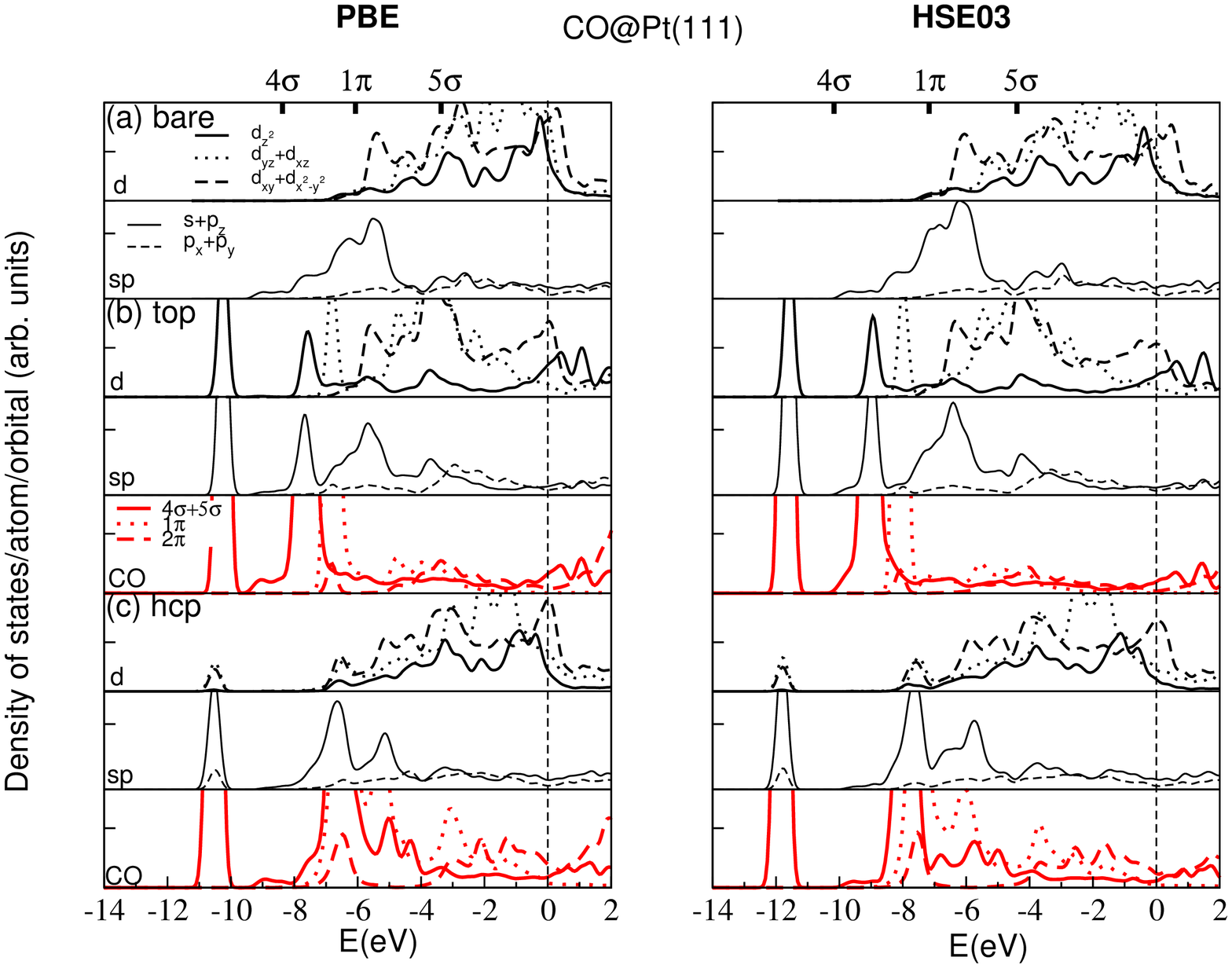}
\\Fig. 3
\end{center}
\caption{(Color online) Orbital resolved electronic density of states (DOS) for the
topmost Pt layer of the bare surface (a) and for CO
adsorbed at the (b) top and hcp (c) fcc site. See Caption of Fig.~\ref{dosCu} for details.}\label{dosPt}

\end{figure}

 In order to gain insight on the effect of hybrid functionals on the electronic properties,
  we show the orbital resolved
 electronic density of states (DOS) for the Cu, Rh and Pt
 surfaces in Fig.~\ref{dosCu}, Fig.~\ref{dosRh}, and Fig.~\ref{dosPt} respectively.
 For each Figure, we show the DOS for
 the bare surface layer [(a), panels  d, sp], for  CO adsorbed
  on the  top [(b), panels d, sp, CO], and hollow site
 [(c), panels d, sp, CO] as obtained
 using PBE (left column) and HSE03 (right column). Here d, sp, CO indicate the projection onto $d$, $sp$ metal states,
 and CO molecular orbitals respectively. In panel (a)-d, we show the one-electron energies of the $4\sigma$, $1\pi$,
  $5\sigma$
  molecular levels aligned with the Fermi level using bold tick marks.\cite{footnote1}

 Since there are no principal differences between the PBE0 and HSE03 DOS,   only
 results for the latter are shown. We caution the reader that the following analysis  based on the atom-projected DOS
  can only  give qualitative insight on variations of the
 chemisorption energies upon changing the adsorption sites,
 the metal surface and  the exchange-correlation functional.
 A more powerful and quantitative analysis can be
 done by using the concept of \emph{surface group orbitals}  introduced
 by R.A. van Santen,\cite{Santen1,Santen2,Santen3} but this is
 beyond the purpose of this paper.

 We recall the basic features of the interaction of the CO molecule with the metal
 surface,\cite{Kresse1,Bagus,Hoffmann,Santen4} and we first focus on the
   PBE DOS for the Cu case (Fig.~\ref{dosCu},
  left column). From panel (a)-d and (a)-sp, we note that  the
  Cu $d$ band is \emph{almost} completely filled,
  it is centered around $\epsilon_{d} \sim-$2.4
  eV ($d$ band center of gravity) with $\delta \epsilon_{d}$=1.52 eV
  ($d$ band width);\cite{footnote3}
  the $sp$ band extends over a larger energy
  range below the Fermi energy, but it is strongly peaked around $-$5
  eV;
   the $5\sigma$ level is located at $\sim-$2.70 eV,
    and the $2\pi^{*}$ level   is positioned at $\sim$ 4.20 eV above the Fermi energy.

  Let us now consider  the interaction of the metal states
  with the CO $\sigma$ orbitals in the top configuration [Fig.~\ref{dosCu} (b)].
 The metal $d$ bands broaden and  shift down to lower energies due to the
 interaction with the CO molecule [compare panels (b)-d with (a)-d]. In particular, the
 $d_{z^{2}}$ DOS is strongly modified.
  From panel (b)-CO we see that the $4\sigma$ and $5\sigma$ orbitals are
  shifted to lower energy (two main peaks at $-$10 and $-$7 eV  with
  continuous line).
  Comparing panels in (b), it is clear that  a
  strong interaction between  $d_{z^{2}}$ and $sp_{z}$ metal states and the $\sigma$  molecular
  states   takes place giving
  rise to a bonding contribution
 below the Fermi level (below $\sim -$ 6 eV) and non-bonding and
 antibonding contributions partly even above the Fermi
 level (d$_{\sigma}$ band).\cite{Fohlish,Santen4}
 We recall that, since  almost fully occupied states are interacting, the
 interaction would be only a Pauli-like repulsion, if the
 antibonding $\sigma-d_{z^{2}}$ hybrid states were not pushed
 above the Fermi level (relief of Pauli repulsion).\cite{Pauli}
  The corresponding depletion of $5\sigma$ states
 (donation from CO to the metal) is in accordance with the Blyholder model.

 Let us now consider the $\pi$ orbitals.
 From panel (b)-CO, we note that they are
 shifted to lower energy due to the interaction with $d_{yz}$,
 $d_{xz}$ and $sp$ metal states (bonding contribution).
 The $1\pi$ states are found at $\sim-$7 eV (dotted line), and a small  peak  of $2\pi^{*}$
  symmetry at the position of the $1\pi$ orbital is visible and related  to the $1\pi-2\pi^{*}$
  hybridization.\cite{Fohlish} Above $-$6 eV, a non-bonding and antibonding $1\pi-$derived
  band  develops (d$_{\pi}$ band)\cite{Fohlish} which, however, remains  below the Fermi energy
  (Pauli-like repulsion). The bonding  states due to the interaction between the
  $2\pi^{*}$ orbital    and the $d$ band are found around $-$4 eV
  (again $d_{\pi}$ complex)\cite{Fohlish} and antibonding
  interactions well above the Fermi energy (not shown in the Figure). The partial occupation of
  the 2$\pi^{*}$ molecular orbital corresponds
  to a back-donation of electrons from the substrate to the originally empty  2$\pi^{*}$  molecular
  orbital, in accordance with the Blyholder model.

 For the hcp site, there are some differences. The $d_{z^{2}}$ orbital is only
 marginally affected upon adsorption, whereas the $d_{yz}$
 ($d_{xz}$) and  $d_{xy}$ ($d_{x^{2}-y^{2}}$) bands are broadened
 and shifted to lower energy [compare panel (c)-d with (a)-d].
 The interaction is mainly with the $\pi$ orbitals:
 the peaks with
 dashed and dotted lines above $-$6 eV in panel (c)-CO have increased compared to
 the corresponding ones in panel (b)-CO.
 In accordance with the arguments of Ref.~\onlinecite{Santen4},
 the interaction between the metal $s$ and $p$ states and the CO $\pi$ states
 is also enhanced, as reflected by a downshift of the metal $sp$ states. Note that
 three metal atoms are affected by adsorption in the hollow site as compared to one
 metal atom for the top site.
 In summary, the $\pi-d$ and $\pi-sp$ interaction is larger for the hcp site than for atop adsorption.

 At this point, before considering the HSE03 DOS,
 we stress again that CO chemisorption to a
 transition-metal surface involves interaction of the molecular frontier orbitals
 and the broad $s$ and narrow $d$ band.
 The question whether the single-atom or high
 coordination site is favoured is clearly the result of a subtle
 balance between  \emph{single-atom} favouring interactions with the
 $5\sigma$ molecular states and \emph{high-coordination} favouring
 interactions with the $2\pi^{*}$ molecular states.
 For CO on Cu(111)  the balance between \emph{single-atom} and \emph{high-coordination} directing interactions
 is such that the hollow site adsorption is favoured for the PBE functional.

  We now turn our attention to Fig.~\ref{dosCu} right column.
  For the bare surface [(a)-d],
  we see that the $d$ states are shifted to lower energy by $\sim$ 1 eV with respect to
  the Fermi level compared to the PBE DOS ($\epsilon_{d}$=$-$3.6
  eV), but
  the overall spectral shape and the bandwidth remain
  almost unchanged. The downward shift of
  the HSE03 $d$ bands can be understood as a result of the
  reduction of  the self-interaction (SIC) within the $d$ shell in the
  hybrid-functional formalism. It is also
  clearly seen that the occupied (unoccupied) molecular levels are shifted downwards (upwards)
  with respect to the Fermi energy: the $5\sigma$ level is now located  at
  $-$4.30 eV and the $2\pi^{*}$ level is at 4.50 eV. The
  change of the energy positions of the  non-interacting
  molecular levels with respect to the Fermi energy is mainly related
  to the increased HOMO-LUMO gap in the free molecule, but also partly caused by the reduced
  work function for the HSE03 functional (6.10 eV compared to 6.30 eV for PBE).
 The effect of the inclusion of part of the exact exchange on the broad  $sp$ band is different.
 As a matter of fact, the $sp$ band is  restrained to be located at the Fermi level (it is partially occupied),
 and the main effect of the non-local exchange is
 to increase the total bandwidth by 0.70 eV [compare left and right panel (a)-sp].
  The above analysis immediately leads to a
  first conclusion. The down-shift of the  $d$ band and
 \emph{simultaneous} up-shift of the $2\pi^{*}$ levels suggest a
 reduced $2\pi^{*}-d$ interaction: according to second order
 perturbation theory, the larger energy distance between the
 unperturbed energy levels weakens the  interaction.
 Inspection of the calculated DOS confirms this conjecture for HSE03:
 i) the $d_{\pi}$  peaks are much weaker,
 ii)   hybridization between $2\pi^{*}$ and $1\pi$ molecular orbitals decreases
 corresponding to a
 \emph{decrease} of the bonding interaction between the CO $\pi-$states and the metal $d$ states
 according to F\"{o}hlisch.\cite{Fohlish}


 We now summarize our analysis concerning the differences between the PBE and HSE03 functional.
 For the non-interacting fragments (bare surface and CO molecule) the HSE03 functional
 i) pushes the occupied molecular levels
 down in energy, whereas
 the unoccupied molecular orbitals are moved up
 in energy, and ii)  the fully occupied $d$ band is shifted down in energy.
  We have seen that these two \emph{combined} effects 
 generally disfavour the $\pi-metal$
 interaction. It  decreases for both the atop and the hollow site,
 but the destabilization is stronger in the hollow sites,
 in accordance with the observation that the $2\pi^{*}$ interaction is dominant at the hollow sites.
 This is confirmed by the observation that the adsorption energies in Tab.~\ref{TAB3}
 decrease for both atop and hollow sites,  but  the effect is twice as strong
 for the hollow sites (compare the relative variation of the
 adsorption energies from PBE to HSE03, in Tab.\ \ref{TAB3}).
 A quantitative confirmation of this picture stems from the occupation of the
 $2\pi^{*}$ orbital in the hollow site. We recall that the larger   the $2\pi^{*}$ occupation is,
 the stronger is the interaction of $2\pi^{*}$ with the metal states.\cite{Santen4}
 The $2\pi^{*} $occupation is
 0.80 and 0.66 electrons  using PBE and HSE03
 respectively, hence the occupation decreases by $\sim$ 17\% from PBE
 to HSE03.

 Let us now consider  the CO adsorbed on the Rh(111) surface
 in Fig.~\ref{dosRh}. In the following, we mainly concentrate on  the differences between
 the PBE and HSE03 description (left and right panels).
 For the bare surface and HSE03 [panel (a)-d], there is a general downshift
(upshift) of the occupied (unoccupied) part of the $d$ bands with respect to the
Fermi energy. This leads to a small downshift of the   center of
gravity  ($\epsilon_{d}$=$-$1.86 and $-$2.23 eV using PBE and HSE03)
and
 a sizeable larger bandwidth ($\delta \epsilon_{d}$=7.40 and 8.90 eV
using the PBE and  HSE03 functional).
For atop adsorption [panel
(b)-d], we see a
small decrease of the intensity of the corresponding $d_{\pi}$ band
[dotted line in panel (b)-CO]. There is also a small reduction of the
$1\pi$ and $2\pi^*$ hybridization  and of the $2\pi^*-d$
interaction (reduction of the intensity of the corresponding broad
bands with dashed line). As far as the $\sigma-d$ interaction is
concerned, one can see from panel (b)-CO that  the $d_{\sigma}$ band is
slightly less intense at the HSE03 level (reduction of the
interaction strength). For the hcp site, we observe the same trends
as for the top site.

Let us summarize again the differences between PBE and HSE03. As
opposed to Cu, the $d$ band is only slightly shifted downwards, but
posses a much larger bandwidth (1.50 eV). As a result, the strength
of the $\pi-d$ interaction is  only slightly reduced using HSE03:
the
 reduction of the $\pi-d$ interaction caused by the up-shift of the
 $2\pi^{*}$ orbital is counteracted by the larger $d$ bandwidth of the
 metal  using HSE03. This is corroborated by the
 $2\pi^{*}$ occupation for the hollow site, which changes only from 1.05   to 1.01 electrons
  ( \emph{i.e.}  $\sim$ 4 \%) going from PBE to
 HSE03. We recall that for Copper, the variation was about 17 \%,
 \emph{i.e.} much  larger than for Rh.
 Secondly, the $s$ electrons are largely affected by the
 introduction of the non-local exchange. For the bare surface,
  HSE03 reduces the self-interaction  within the $s$ shell
 and moves them  to lower energies. 
 This
 suggest an enhancement of the back-donation due to the interaction of the 2$\pi^{*}$ states
 with antisymmetric linear combinations
 of metal $s$ orbitals, which, we recall, is active only for the hollow site.\cite{Santen4}
 For Rh, the balance between atop- and hollow-directing interactions
 gives still a slight  preference for atop adsorption using HSE03 (cfr. Tab.~\ref{TAB3}).

 Finally we turn our attention to Pt(111).
 Basically we observe the same trends as for Rh using HSE03, a
 down-shift
 of the $d$ band center of gravity  and a larger bandwidth is observed ($\epsilon_{d}$=$-$2.13 and $-$2.53 eV,
 $\delta \epsilon_{d}$=8.29 and  9.32 eV  using PBE and HSE03, respectively).
  The work-function changes only a little
 (5.69 and 5.64 using  PBE and HSE03). 
 For both adsorption sites, there are only small changes in the DOS   due to the introduction of exact
 exchange. In particular, for the hollow site, we note that the peak corresponding to the
 $2\pi^{*}$ and $1\pi$ hybridization has the same intensity for HSE03 and PBE [panel (c)-CO].
 Also the $d_{\pi}$ intensity is almost the same as in the PBE case [panel (c)-CO, dashed line]
 as confirmed by the $2\pi^{*}$
 occupation: it is 1.03 and 0.97  electrons using PBE and HSE03. \emph{i.e.} it decreases by only 6 \% using HSE03.
 Probably, the most pronounced difference between the $4d$ Rh and $5d$ Pt metal is the enhanced $s$ occupation,
  related to the stronger
 binding of the $6s$ electrons (the shell structure requires that the orbitals are filled in the
  order $6s$, $4f$, $5d$, $6p$ placing the $6s$ electrons at
 significantly larger binding energies than $5s$ electrons). This effect is further enhanced by the HSE03 functional.
 As before, this should stabilizes the CO in the hollow site due to
 the increased $\pi-s$ interaction. Indeed now
  the destabilization of the $\pi-d$ interaction in the hollow site is
 not sufficient to  yield the top site as preferred adsorption site.

\section{Discussion}\label{Discussion}

Let us start with a brief discussion of the computational aspects of
the current work. We have shown that periodic slab calculations
using hybrid Hartree-Fock density functionals are perfectly feasible
for metallic systems using a plane wave basis set. We have also
shown that the HSE03 functional, suggested by Heyd, Scuseria and
Ernzerhof, yields practically identical results as the more
conventional PBE0 functional, albeit, at a computational cost that
is reduced by almost a factor ten. This is achieved by replacing the
long range part of the Fock-exchange by its density functional
approximation, leading to a  rapid k-point convergence of the
non-local exchange  and total energies. Thus the HSE03 functional
presents a promising functional for large scale studies of molecules
on surfaces.

Our study concentrated on the chemisorption of the CO molecule on
$d$-metal surfaces, specifically  Cu(111), Rh(111) and Pt(111). The
study has been pursued using  local and generalized gradient density functionals, and
PBE0 and HSE03  hybrid
 Hartree-Fock density functionals. As expected, the  LDA and GGA functionals give the
wrong site preference for Cu, Rh and Pt. In contrast, the PBE0 and HSE03 functionals reduce this tendency,
predicting the correct site order for CO on Cu(111) and Rh(111). In
both cases, the fcc and hcp sites are destabilized by roughly 150
meV compared to the top site. Unfortunately the HSE03 and PBE0 functionals
do not work so well for Pt(111), where the destabilization is only 50
meV for the fcc site and 80 meV for the hcp site, which is not
sufficient to yield the correct site preference. In
both Rh and Pt, we have  made significant efforts applying for
instance different PAW sets and parameters (not all have been
discussed in detail) to make certain that the present numbers are
essentially converged within the theoretical framework.

The wrong site order for Pt is not the only unsatisfactory aspect of
our study; results for the energetics are also largely
disappointing. It is well accepted that gradient corrected
functionals have a tendency to overestimate adsorption energies on
metal surfaces.\cite{CO_Pt,rPBE} One would have hoped that admixing
a certain fraction of the exact non-local exchange lifts this
deficiency, but this hope is not fulfilled by the PBE0 and HSE03
functionals. In fact, Cu is the only case where the hybrid
functionals improve the overall energetics. This is related to the
upshift of the empty CO $2\pi^{*}$ orbital and a \emph{simultaneous}
downshift of the filled Cu 3$d$ states, with both effects reducing,
in concert, the $2\pi^{*}-d$ interaction; thus the top site becomes
preferred.

For Rh and Pt, the $d$ band is restrained to stay at the Fermi-level,
and for the transition metals the main effect of the inclusion of
non-local exchange is an increase of the $d$ bandwidth. This
increase of the $d$ bandwidth counteracts the reduced interaction
caused by the upshift of the CO $2\pi^{*}$ orbital. One therefore
observes that the interaction energies generally increase from PBE,
over HSE03 to PBE0, with the last one yielding the largest $d$ band
width and the largest CO-metal interaction energies. In a ball-pack,
the increased metal bandwidth caused by the non-local exchange is the
main origin of problems: it partially restores the CO $2\pi^{*}-d$
interaction that we had aimed to reduce by means of the hybrid
functional. This counterbalance works efficiently for Pt, which has the largest $d$  bandwidth
  and the largest interaction  matrix
elements between molecule and metal states.\cite{HammerNorsk}
We also qualitatively argued that further contributions in favor of restoring the back-donative interaction
may come from an enhanced interaction of $2\pi^{*}$ states at the hollow site with \emph{antisymmetric} combination
of $s$ metal states,\cite{Santen4} but our analysis is not able to quantify and separate $s$ and $d$  contributions.

 Unfortunately, there are  reasons to believe that
the inclusion of a significant fraction of the non-local exchange
and the concomitant increase of the $d$ bandwidth in transition metals
is the wrong physics. We have already commented on this issue in our
recent work:\cite{Paier2} the analogy between $GW$ and hybrid
functionals suggests that the amount of non-local exchange should
be chosen system dependent, applying more Hartree-Fock like exchange
in exchange dominated systems such as molecules and large gap insulators.
In metals, on the other hand, the non-local exchange term in
$GW$ is almost entirely screened by the other electrons, so that the
Coulomb hole term--- corresponding to a local potential ---becomes
dominant. In metals, the proper description thus involves only a
very weak screened-exchange interaction and the semi-local density
functionals approximation should do a  perfectly adequate job. The same
conclusion is reached using the adiabatic connection fluctuation
dissipation theorem (AC-FD).\cite{PBE03}  In a forthcoming
paper,\cite{ASGK} we will present results of hybrid-funtional
calculations of CO adsorption extended to other systems
representative of $4d$ and $5d$ metal surfaces, including the B3LYP
functional. Unfortunately, also in this case, the results are
generally discomforting, suggesting that previous reports on the
successful prediction of adsorption energies using B3LYP have to be
considered with suspicion.

We are thus left with the intriguing problem, how to treat two
disparate systems using the same unified theoretical footing. It is difficult to imagine that a  hybrid
functional with a fixed amount of non-local exchange is  going
to do the job. On passing, we also reiterate
another  result for hybrid functionals from Ref.~\onlinecite{Paier2}: the
exchange splitting in transition metals is significantly
overestimated, resulting in too small atomization energies (found
here as well for Pt) and a large overestimation of the magnetic
moment in itinerant magnetic transition metals.

\section{Conclusions}\label{conclu}
We conclude that, although hybrid functional calculations for metals
and metal surfaces are perfectly feasible, the results are by no
means entirely satisfactory. The agreement with experiment is
improved for CO on Cu(111), but  the results are only marginally
improved for CO on Rh(111) (correct site order but much too large
adsorption energies) and hardly improved for CO on Pt(111) (wrong
site order and too large adsorption energies). We have argued that
this failure is related to the inclusion of non-local exchange in
the metal slab which results in an incorrect description of the
metal band width.

\section{Acknowledgments}
This work was supported by the Austrian {\em Fonds zur F\"orderung
der wissenschaftlichen Forschung}.

\newpage

\begin{table}[!hbp]
\caption{Lattice constants $a_{0}$ (\AA), bulk moduli $B_{0}$
(GPa), cohesive energies $E_{coh}$ (eV) of bulk Cu, Rh and Pt
obtained from LDA, GGA (PBE), PBE0 and HSE03 calculations. Results for HSE03
are based on calculations employing a 'reduced'
 $12\times12\times12$  k-point grid, those for LDA, GGA (PBE) and PBE0 a full
 $12\times12\times12$  grid, i.e. without downsampling (see text for
details). Relative errors (\%) with respect to experiment are shown
in round brackets.} \label{TAB1}
\vspace{1cm}
\begin{tabular}{cccc}
\hline
  & a$_{0}$(\AA) & $B_{0}$ (GPa) &$E_{coh}$(eV)\\
\hline
\multicolumn{4}{c}{Cu} \\
LDA & 3.524  ($-2.2$~\%) & 184 ($+29.0$ \%)& 4.498 ($+$28.9 \%)       \\
PBE & 3.635 ($+$0.9~\%)   & 136 ($-4.2$ \%)& 3.484 ($-0.2$ \%)    \\
PBE0 & 3.636 ($+$0.9~\%) & 130 ($-8.4$ \%)  &  3.046 ($-12.7$ \%) \\
HSE03 &3.640 ($+$1.0~\%)  & 135 ($-4.9$ \%)& 3.066   ($-12.1$ \%) \\
B3LYP (Ref.\onlinecite{Neef}) & 3.700 ($+$2.7 \%) & 117 ($-17.6$ \%)&2.892 ($-17.1$ \%)\\
Exp & 3.603 & 142&3.49 \\

\multicolumn{4}{c}{Rh} \\
LDA & 3.752 ($-1.2$ \%) & 318 ($+$18.2 \%)& 7.382 ($+$28.4 \%)  \\
PBE &  3.823 ($+$0.6 \%)   & 254 ($-5.6$ \%) &  5.724 ($-0.4$ \%) \\
PBE0 & 3.785 ($-0.3$ \%)  & 291  ($+$8.2 \%) &  4.205 ($-26.9$ \%) \\
HSE03 &3.783 ($-0.4$ \%)  &  305 ($+$13.4 \% ) &  4.441 ($-22.8$ \%) \\
Exp & 3.798  & 269 & 5.75\\

\multicolumn{4}{c}{Pt} \\
LDA  & 3.905 ($-0.4$ \%) & 306 ($+$10.0 \%)    & 7.076 ($+$20.9 \%) \\
PBE & 3.965 ($+$1.2 \%)    &277 ($-0.4$ \%)    &   5.668  ($-3.1$ \%)   \\
PBE0 & 3.932 ($+$0.3 \%)   &  274 ($-1.4$ \%)  &   4.648 ($-20.5 $ \%)   \\
HSE03 & 3.932 ($+$0.3 \%  )   & 275($-1.1$ \%) &  4.900 ($-16.2 $ \%)    \\
B3LYP (Ref.\onlinecite{Doll}) & 4.05 ($+$3.3 \%) & 234 ($-15.8$\%) & 3.755 ($-35.8$ \%) \\

Exp & 3.920    & 278  & 5.85 \\
\hline

\end{tabular}
\end{table}
 \newpage
\clearpage

\begin{table}[!hbp]
\caption{Relevant equilibrium properties of the bare surfaces using the LDA, GGA (PBE), PBE0, and HSE03 functionals. $d_{12}$ is the interlayer distance
 (in \AA)
between the first and second   topmost layers.
Numbers in round brackets refer to the change (in \%)  relative to the theoretical
interlayer bulk distance ($a_{0}/\sqrt 3$). $E_{surf}$ is the average
surface energy of the relaxed and unrelaxed side of the
slab, normalized to the (111) surface unit cell. In this and the  following tables,
calculations for the 4 layer thick slab using $6\times6\times1$ (structural properties) and
  $8\times8\times1$  (energetics) k-point grids are reported. All calculations were
performed using a  $c(2\times4)$ surface unit cell.}
\label{TAB2}
\vspace{1cm}
\begin{tabular}{ccc}
\hline
  & $d_{12}$(\AA)&  $E_{surf}$(eV/(111) unit cell)\\
\hline
\multicolumn{3}{c}{Cu} \\

 LDA   &   2.01  ($-0.9$~\%)  &    0.588    \\
 PBE   &   2.08  ($-0.9$~\%)  &    0.455    \\
PBE0  &   2.06   ($-1.8$~\%)  &    0.465    \\
HSE03 &   2.08   ($-0.9$~\%)  &    0.453    \\

B3LYP  (Ref.\onlinecite{Neef}) & 2.11 ($-1.2$\%) & 0.500  \\

\multicolumn{3}{c}{Rh} \\

LDA  &   2.13   ($-1.6$~\%)  &  1.006  \\
PBE &    2.17   ($-1.6$~\%) &   0.814  \\
PBE0 &   2.17   ($-0.6$~\%) &    0.849  \\
HSE03 &  2.16   ($-1.0$~\%)  &    0.845  \\


\multicolumn{3}{c}{Pt} \\


LDA &  2.26    ($+$0.4 \%) &     0.813  \\
PBE & 2.30    ($+$0.7 \%)  &     0.618  \\
PBE0 & 2.30  ($+$1.3 \%)  &      0.644  \\
HSE03 & 2.30  ($+$1.3 \%) &      0.672  \\
B3LYP (Ref.\onlinecite{Doll})& 2.39 (+2.1~\%) &  0.517  \\
\hline

\end{tabular}
\end{table}
 \newpage
\clearpage

\begin{table}[!hbp]
\caption{Relevant structural parameters and energetics of CO on the Cu(111) surface (top, fcc, and hcp sites)
 using the LDA, GGA (PBE), PBE0, and HSE03 functionals (all the distance in ~\AA).
  d$_{C-O}$ is the vertical distance between C and O, and in parenthesis the relative
  variation with respect to the theoretical value of
  the free CO molecule (\%) is given; $d_{C-X}$  is defined
  as the \emph{minimum} heigth difference between the $z$ coordinate of
   the C  and   metal atom  directly
  involved in the C-X bonds; buckling $b$ is the distance between the outermost and the innermost metal atom
 in the first layer; $d_{12}$ is the mean change (\%) of the distance between the first and second layer
 with respect to the theoretical value of the bare unrelaxed surface; $E_{ads}$ is the adsorption energy in eV.
 The preferred site (and corresponding adsorption energy)
 is written in boldface.}
\label{TAB3}
\vspace{1cm}
\begin{tabular}{ccccccc}
\hline
Method& Site& $d_{C-O}$(\AA)  & $d_{C-X}$(\AA)& $b$(\AA)&$d_{12}$(\%)&$E_{ads}$(eV)\\
\hline
\multicolumn{7}{c}{Cu} \\

LDA   & top        &  1.149  ($+$1.2\%)& 1.798 & 0.088 &$-$1.0\%  & $-1.286$       \\
      & {\bf fcc } &  1.174  ($+$3.5\%)& 1.355 & 0.094 &$+$0.5  \%  & ${\bf-1.660}$  \\
      & hcp       &   1.174  ($+$3.5\%)& 1.360 & 0.068 &$+$0.4  \%  & $-1.642$       \\

PBE & top       &  1.158 ($+$1.3 \%)&1.844 & 0.124&$-1.2$  \%  & $-0.709$   \\
      & {\bf fcc } &  1.183 ($+$3.5  \%) &1.395 & 0.109&+0.5 \%  & ${\bf -0.874}$  \\
      & hcp       & 1.182  ($+$3.4 \%)& 1.399 & 0.082&+0.6 \%  & $-0.862$   \\

PBE0& {\bf top}  & 1.144 ($+$0.9 \%)& 1.856 & 0.138& $-1.7$  \%& ${\bf -0.606}$ \\
      & fcc      & 1.163 ($+$2.6 \%)&1.425 & 0.135&$+$0.7  \%&$-0.579$ \\
      & hcp      & 1.161 ($+$2.5 \%)&1.450 & 0.105&$+$0.7  \%& $-0.565$\\

HSE03 &{\bf top}&1.142 ($+$1.0 \%) &1.864 & 0.146 & $-1.7$  \%& ${\bf -0.561}$   \\
       &fcc  &1.160 ($+$2.6 \%)&1.417 & 0.143 & $+$0.7  \%&  $-0.555$ \\
       &hcp  &1.158 ($+$2.4 \%)  &1.413 & 0.099 & $+$1.2  \%& $-0.535$\\
\hline
\end{tabular}
\end{table}
 \newpage
\clearpage

\begin{table}[!hbp]
\caption{Relevant structural parameters and energetics of CO on the Rh (111) surface (top, fcc, and hcp sites)
 using the LDA, GGA (PBE), PBE0, and  HSE03 functionals. See caption \ref{TAB3} for details.
}
\label{TAB4}
\vspace{1cm}
\begin{tabular}{ccccccc}
\hline
Method& Site& $d_{C-O}$(\AA)  & $d_{C-X}$(\AA)& $b$(\AA)&$d_{12}$(\%)&$E_{ads}$(eV)\\
\hline
\multicolumn{7}{c}{Rh} \\
 LDA  & top  & 1.155 ($+$1.8\%) & 1.810 & 0.167 & $+$0.0\%  &    $-2.480$    \\
      & fcc  & 1.185 ($+$4.5\%) & 1.350 & 0.083 &  $+$1.0\%  &    $-2.733$  \\
      & {\bf hcp}  & 1.187 ($+$4.6\%) & 1.330 &0.083 & $+$0.8\%  &   ${\bf-2.801}$  \\

PBE & top  & 1.165 ($+$1.9 \%) & 1.827 & 0.213  &$+$0.3 \%  &    $-1.870$    \\
      & fcc  & 1.195 ($+$4.5 \%) & 1.379 & 0.082 &$+$ 0.9 \%  &    $-1.906$  \\
      & {\bf hcp}  & 1.197 ($+$4.7 \%) & 1.348 & 0.096 &$+$1.3 \%  &   ${\bf -1.969}$  \\

PBE0& {\bf top}  & 1.149 ($+$1.4 \%) & 1.834 & 0.234  & $+$2.5 \%      & ${\bf -2.109}$ \\
      & fcc  & 1.185 ($+$4.6 \%)& 1.330 & 0.048  & $+$2.8 \%          &  $-2.024$  \\
      & hcp  & 1.185 ($+$4.6 \%)& 1.344 & 0.118  & $+$4.3 \%          &  $-2.104$   \\

HSE03 & {\bf top}    & 1.152 ($+$1.8 \%)& 1.811    & 0.172  &$+$0.4\%   &  ${\bf -2.012}$\\
      &   fcc        & 1.193 ($+$5.5 \%)& 1.351    & 0.063  &$+$1.9\%   &  $-1.913$  \\
      &   hcp        & 1.191 ($+$5.3 \%)& 1.342   &  0.100  &$+$2.2\%   &  $-1.996$  \\
\hline
\end{tabular}
\end{table}
 \newpage
\clearpage

\begin{table}[!hbp]
\caption{Relevant structural parameters and energetics of CO on the Pt (111) surface (top, fcc, and hcp sites)
 using the LDA, GGA (PBE), PBE0, and HSE03 functionals. See caption \ref{TAB3} for details.}
\label{TAB5}
\vspace{1cm}
\begin{tabular}{ccccccc}
\hline
Method& Site& $d_{C-O}$(\AA)  & $d_{C-X}$(\AA)& $b$(\AA)&$d_{12}$(\%)&$E_{ads}$(eV)\\
\hline
\multicolumn{7}{c}{Pt} \\

LDA &  top  & 1.149 ($+$1.2\%) & 1.827  & 0.186   & $+$0.5\%  &   $-2.251$    \\
      & {\bf fcc}  &  1.184 ($+$4.3\%)  & 1.329 & 0.130  &  $+$2.5\%  &  ${\bf -2.601}$  \\
      & hcp   &  1.183($+$4.2\%)  & 1.314 & 0.132  & $+$2.0\%  &  $-2.576$   \\

PBE & top  & 1.158 ($+$1.3 \%) & 1.839  &  0.227  & $+$0.5  \%  &   $-1.659$    \\
      & {\bf fcc}  & 1.194 ($+$4.4 \%)  & 1.329  &  0.132 &  $+$2.3 \%  &  ${\bf -1.816}$  \\
      & hcp   & 1.194 ($+$4.4 \%)  & 1.324  & 0.149  & $+$2.4 \%  &  $-1.750$   \\

PBE0  & top  & 1.142   ($+$0.8 \%)   & 1.818   & 0.237     & $+$0.8    \%     &  $-1.941$    \\
        & {\bf fcc}  & 1.177   ($+$3.9 \%)   & 1.304   & 0.215    &  $+$3.2   \%     & ${\bf  -1.997}$  \\
        & hcp  & 1.180   ($+$4.1   \%)  & 1.291   & 0.226     &  $+$2.6   \%    &   $-1.944$   \\

HSE03 &  top   & 1.143 ($+$1.1  \%) & 1.821 & 0.200& $+$0.5 \%  &   $-1.793$ \\
        &  {\bf fcc}   & 1.177  ($+$4.1 \%) & 1.320 & 0.177&  $+$3.3  \%  &  ${\bf-1.862}$  \\
        &  hcp   & 1.177  ($+$4.1 \%) & 1.330 & 0.177 & $+$2.2\%  &  $-1.808$ \\

\hline
\end{tabular}
\end{table}
 \newpage
\clearpage
\newpage

\begin{table}[!hbp]
\caption{Dependence of the  site order on the  sampling of the
surface Brillouin zone using hybrid functionals and the PBE functional.
 Energies are referenced to the top site. PBE0   results are
 evaluated using the "full" k-point grid; HSE03 results are evaluated using the "reduced" as well as  the
 "full" grid (numbers in round
 brackets) for the non-local exchange, see text for details.}
\label{TAB2.5}
\vspace{1cm}
\begin{tabular}{ccccccc}
\hline

&\multicolumn{2}{c}{PBE0}& \multicolumn{2}{c}{HSE03}&\multicolumn{2}{c}{PBE}\\

&fcc&hcp&fcc&hcp&fcc&hcp\\
\hline
\multicolumn{7}{c}{Cu}\\

  $4\times4\times1$  &   $\;\:$ 0.095  & $\;\:\;$0.097& $\;\:\;$0.099 ($\;\:\;$0.086)&$\;\:\;$0.106 ($\;\:\;$0.094)&$-$0.097 &$-$0.089\\

  $6\times6\times1$ & $-$0.029& $-$0.004  &$-$0.019 ($-$0.024) &$\;\:\;$0.010 ($\;\:\;$0.003)&$-$0.175& $-$0.158 \\

  $8\times8\times1$ & $\;\:\:$0.027  & $\;\:\;$0.041 & $\;\:\;$0.006 ($\;\:\:$0.006) &$\;\:\;$0.026 ($\;\:\;$0.028)&$-0.165$&$-0.153$ \\
 \hline
\multicolumn{7}{c}{Rh}\\
  $4\times4\times1$  &$\;\:\;$0.099&$\;\:\;$0.000  &$\;\:\;$0.079  ($\;\:\;$0.081) &$-0.006$ ($-0.022$) & $-0.040$ & $-0.105$\\

 $6\times6\times1$  & $\;\:\;$0.118& $-$0.007&$\;\:\;$0.076 ($\;\:\;$0.074) & $\;\:\;$0.005 ($-$0.003) &$-$0.067& $-$0.130\\

  $8\times8\times1$ & $\;\:\;$0.085& $\;\:\;$0.005&$\;\:\;$0.089 ($\;\:\;$0.094) & $\;\:\;$0.016 ($\;\:\;$0.048)& $-$0.061 & $-$0.131\\
 \hline
\multicolumn{7}{c}{Pt}\\
   $4\times4\times1$  & $\;\:\;$0.047    & $-0.027$ & $-$0.091 ($-$0.112) & $-$0.045 ($-$0.062)& $-$0.127 & $-$0.106\\
  $6\times6\times1$ & $-$0.073& $\;\:\;$0.067 & $-$0.062  ($-$0.070)& $-$0.011 ($-$0.014) &  $-$0.118& $-$0.094\\
  $8\times8\times1$ & $-$0.056  & $-$0.003 &  $-$0.069  ($-$0.069) &  $-$0.015  ($-$0.015) & $-$0.121 &  $-$0.095\\
\hline

\end{tabular}
\end{table}
 \newpage
\clearpage

\vspace{1cm}
\newpage
\newpage

\vspace{1cm}
\newpage

\vspace{1cm}
\newpage


\begin{thebibliography}{}
\bibitem{footnote1} The electronic molecular energies
  of the free CO molecule calculated in a
   large cube have the vacuum level as zero energy level. By adding
    the work function of the metal surface to the free molecular energy
    levels, the Fermi energy of the metal becomes the zero energy level.
\bibitem{footnote3} The bandwidth and the center of gravity have
   been calculated isolating the $d$ band intensity with a
   cutoff point  chosen to lie at the point of maximum curvature in the upper leading shoulder of
   the DOS spectra
  in order to exclude spurious states   projected
  out from the next $d$ shell. This
 cannot be done in a fully unambiguous way. In any case, reasonable shifts of the cutoff
 point to higher energy
 affect the $d$ band parameters in a negligible way.
\end{thebibliography}
\end{document}